\documentclass{article}
\usepackage{graphicx}
\usepackage{latexsym}

\begin{document}

\title{Superparamagnetism and Monte Carlo simulations}
\maketitle

\author{D. Serantes and D. Baldomir}

\subparagraph{}
Departamento de F\'{\i}sica Aplicada and Instituto de Investigaci\'ons Tecnol\'oxicas, Universidade de Santiago de Compostela, Santiago de Compostela E-15782, Galiza, Spain


\begin{abstract}
We revise the basic concepts beneath the idea of \textit{superparamagnetism} and the suitability of Monte Carlo (MC) simulations to study superparamagnetic (SPM) properties. Starting with the description of the characteristic features of the single-domain SPM entities, their general magnetic-field and temperature-dependent magnetic properties are discussed. Then, the use of a MC technique for studying SPM properties is presented, starting with a general approach to MC methods and introducing the Metropolis algorithm as an adequate tool for reproducing SPM features. Special attention is paid to the role played by the \textit{computational time} MC steps on the simulations.
\end{abstract}

\section{Introduction}
Nanosized magnetic materials exhibit a rich variety of magnetic phenomena in comparison with the bulk counterparts, what gives place to a novel range of applications of great importance to improve daily human activities as high-density magnetic recording storage or biomedical applications\cite{herzer_jmmm2005}. The origin of these special magnetic properties is found on the reduced dimensionality\cite{guimaraes}: when the size of the material reaches the order of nanometers the influence of the surface atoms becomes very comparable (or even higher) than the bulk contribution, the defects due to the broken symmetry may be of significant importance, and other physical effects may also become very relevant when the size reaches the order of characteristic length scales of the material (as for example the domain size). The properties observed in such reduced dimensions are strongly sensitive to small size, shape, and composition dependence, what defines the different magnetic structures (nanoparticles, nanoparticle arrays, nanowires, thin films, etc) as forming specific research fields with differentiated proper characteristics\cite{skomsky_jpcm2003}.

In the first part of this article we revise one of the most remarkable magnetic properties that arises in these reduced dimensions, the so-called \textit{superparamagnetism}. Superparamagnetic (SPM) phenomena is the paramagnetic-like temperature dependence that occurs in single-domain magnetic entities above a characteristic threshold named \textit{blocking temperature}, and its special features are determined by the complex interplay between the magnetic parameters ruling the system (magnetic moment, anisotropy, applied field, etc). Understanding the SPM properties of nanosized systems is of primordial importance both for the basic theoretical knowledge \cite{dormann_fiorani_tronc} and for specific-designed applications (as for example the increase of the storage information capacity of hard drives \cite{skumryev_nature2003}, or the development of well-controlled biomedical applications \cite{tartaj_jpd2003}). Because of this a big effort has been devoted to the study of SPM systems in the last years, aimed to understand its underlying physical mechanisms. 

However, the investigation of SPM properties is a complex task due to its strong dependence on several uncontrolled parameters, which mask the physical origin of the magnetic behaviour and hence makes very difficult to achieve a precise characterization. The high parameter-dispersion degree found in real systems, arising from the large dispersion of parameters (particle size, anisotropy, shape) and uncontrolled interparticle interactions, results in a complex physical problem non-solvable by analytical methods. To investigate such complex scenario it is very common the use of computational techniques, which allow a precise control of the physical parameters governing the system: by means of a computational technique it is possible to set up ideal systems (e.g. monodisperse size and/or anisotropy) specially designed to ease the comprehension of the physical mechanisms governing the system. In the second part of this work we introduce the basic characteristics of a a \textit{Monte Carlo} (MC) method based on the \textit{Metropolis algorithm} to undertake the study of SPM properties.

\section{Superparamagnetism}
\label{sec:Superparamagnetism}

The term \textit{superparamagnetism} refers to the magnetic phenomena observed in fine magnetic particle systems exhibiting close similarities to atomic paramagnetism. Basically, single-domain magnetic nanoparticles can be characterized by their large total magnetic \textit{supermoment}, which exhibits a paramagnetic-like decay of the magnetization with temperature above a characteristic threshold named \textit{blocking temperature}, $T_{B}$. This particular temperature, as difference to the \textit{Curie} temperature $T_{C}$, is extremely dependent on the experimental observational time-window and this characteristic gives place to a complex theoretical frame with especial experimental features. In what follows we briefly introduce the conditions for the existence of superparamagnetism and its basic characteristics. For more information in this topic see for example the reviews by D. Kechrakos\cite{review-kechrakos_JNN}, O. Petracic\cite{review-petracic}, M. Knobel et al.\cite{review-knobel_JNN}, J.L. Dormann et al.\cite{dormann_fiorani_tronc}.

\subsection{Single-domain particles}
\label{sec:SingleDomainParticles}

\begin{figure}[ht!]
\begin{center} 
\includegraphics[draft=false,width=0.50\textwidth]{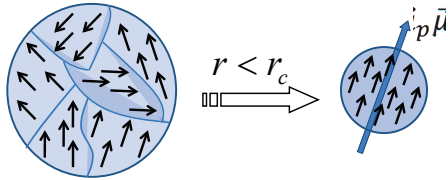}
\end{center} 
\caption{Scheme illustrating the transition from the multi-domain configuration to the single-domain one upon size reduction.}\label{FIG.1-monodomain}
\end{figure}

SPM phenomena is observed upon reduction of the size of ferromagnetic (FM) materials\cite{ferromagnetic} down to the single-domain range: in a FM magnetic material, multiple magnetic domains exist as a result of the balance between the exchange interaction energy favouring the parallel alignment of neighbouring atomic moments (thereby forming magnetic domains), and the magnetostatic interaction energy forcing their breaking into smaller domains with tendency to antiparallel orientation. The domain size is determined by the relative counterbalance between both energies. With decreasing size of the magnetic system, there is a critical value ($r_{c}$ for the radius of a spherical particle) below which the magnetostatic energy no longer allows for the breaking of the system into smaller domains and so the system is composed of a \textit{single domain}, as illustrated in {\sc Figure}~\ref{FIG.1-monodomain}. Typical values are $r_{c}\approx{15}$ \textit{nm} for Fe and $r_{c}\approx{35}$ \textit{nm} for Co \cite{dormann_fiorani_tronc}.

Assuming coherent rotation of the domain atomic moments, the particle is therefore characterized by its total magnetic \textit{supermoment}, $\vec{\mu}_{p}$, resulting from the total magnetization of the particle. In first approximation, considering uniform magnetization (so neglecting surface effects) it can be described as proportional to the particle volume $V$ and saturation magnetization $M_{S}$ as

\begin{equation}\label{eq_mu}
\left|{\vec{\mu}_{p}}\right|=M_{S}V
\end{equation}

As mentioned above, the SPM response in magnetic nanoparticles is observed above the so-called blocking temperature $T_{B}$, a proper feature of SPM systems that differentiates them from atomic paramagnetism. The origin of $T_{B}$ relays on the magnetic anisotropy present in the nanoparticles (in opposition to atomic moments) due to their finite size, which tends to orientate the particle supermoment along some preferential direction. The magnetic anisotropy energy $E_{A}$ found in a magnetic nanoparticle can have different origins (crystal, shape, surface, etc) giving place to very complex scenarios, and so for the sake of simplicity we have focused on the simplest \textit{uniaxial} anisotropy case\cite{anisotropy_sources}. So, from now on we consider the different anisotropy contributions as comprised in an \textit{effective} uniaxial anisotropy term, $\vec{K}_{\mathrm{eff}}$, as illustrated in {\sc Figure}~\ref{FIG.2-NPmoment,eA-wells}(a).

If we consider the magnetic anisotropy to be proportional to the particle volume as $\vec{K}_{\mathrm{eff}}=KV\hat{n}$, with $K$ the \textit{effective} uniaxial anisotropy constant (per unit volume) and $\hat{n}$ the unitary vector describing the easy-magnetization anisotropy direction, then the energy term for the \textit{i}-particle can be written as 

\begin{equation}\label{eq_Ea}
E^{(i)}_{A}=-KV_{i}\left({\frac{{\vec{\mu}_{i}}\cdot{\hat{n}_{i}}}{\left|{\vec{\mu}_{i}}\right|}}\right)^2=-KV_{i}\cos^{2}{\theta}
\end{equation}being $\theta$ the angle between the magnetic supermoment of the particle and the easy anisotropy axis. The moment of the particle has therefore two preferred orientation, equally probable, along the easy-magnetization anisotropy axis direction. Both directions are separated by an energy barrier $E_{B}$ of height $KV$. The energy spectra corresponding to this uniaxial anisotropy energy is illustrated in {\sc Figure}~\ref{FIG.2-NPmoment,eA-wells}(b).

\begin{figure}[ht!]
\begin{center} 
\includegraphics[draft=false,width=0.80\textwidth]{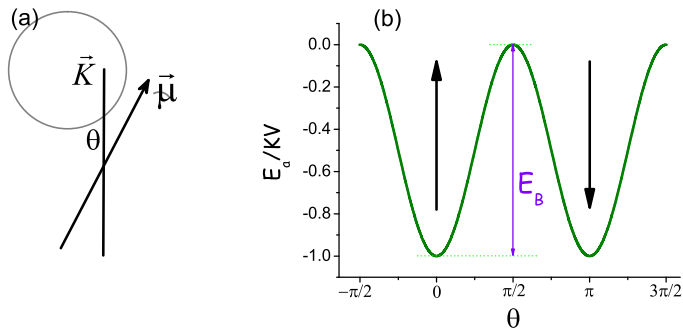}
\end{center} 
\caption{(a) Schematic drawing of the uniaxial magnetic anisotropy $K$ and magnetic supermoment of a single-domain nanoparticle, and (b) the corresponding uniaxial anisotropy wells.}\label{FIG.2-NPmoment,eA-wells}
\end{figure}

The system we have constructed up to now is that of homogeneous magnetic nanoparticles characterized by their size $V$, saturation magnetization $M_{S}$, and uniaxial magnetic anisotropy energy $K$. This very simple scenario describes reasonably well many experimental situations, and so from now on we focus on the magnetic properties of a system of such particles as a function of temperature ($T$) and magnetic field ($\vec{H}$). Real systems are usually characterized by randomness in their spatial distribution and in the easy-axes orientation that strongly influence the properties of the system as determined by interparticle-interactions and applied magnetic field. These analysis are however quite complex and so for the sake of simplicity we consider for the moment a non-interacting particle system (so spatial distribution concerns are avoided) with parallel aligned anisotropy axes, so that the particles are equivalent to each other and the system can be studied under a single-particle perspective.

\subsection{Thermal relaxation and blocking temperature ($T_{B}$)}
\label{sec:ThermalRelaxation}

The influence of the temperature on the magnetic properties of the particles can be easily figured out in the high- and low-T limit cases. At very high T, the thermal energy is much larger than the anisotropy barrier ($k_{B}T >> E_{B}$, with $k_{B}$ the Boltzmann constant) and so the magnetic anisotropy plays a negligible influence on the orientation of the magnetic moments of the particles. In this case a paramagnetic-like dependence of the magnetization with T is expected and the particles are in the \textit{superparamagnetic state} (SPM state). On the contrary, at very low T the particle moment remains confined along the anisotropy direction (local energy minimum) because the thermal energy being unable to switch its orientation out of that minimum ($k_{B}T << E_{B}$). When this happens the particles are said to be in the \textit{blocked state} (BL state).

\begin{figure}[ht!]
\begin{center} 
\includegraphics[draft=false,width=0.65\textwidth]{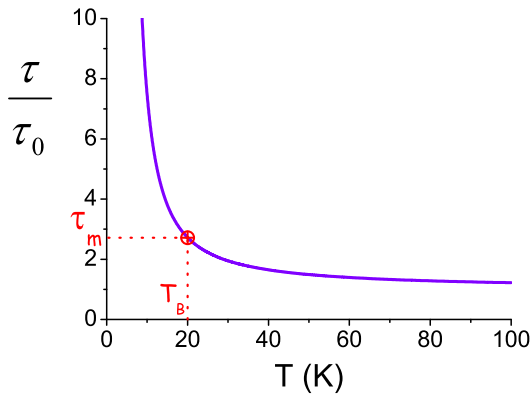}
\end{center} 
\caption{Relaxation time $\tau$ \textit{vs} T, and indication of $T_{B}$ for a certain $\tau_{m}$.}\label{FIG.3-tau_m}
\end{figure}

Thermal energy promotes the fluctuation of the magnetic moments, its influence varying between the rapid motion at very high temperatures and the practically steady state at very low T. Hence, to understand how thermal energy influences the magnetic behavior of the particles it is necessary to understand the dynamics of the particle moments as a function of T. The influence of thermal fluctuation on the orientation of the particles' supermoments was first described by N\'{e}el \cite{neel_1949}, who proposed that the thermal fluctuations could promote the jumping of the magnetic moment of the particles from one anisotropy well to the other, introducing the average time $\tau$ for thermal activation (often called \textit{relaxation time}) over the anisotropy barrier to follow an Arrhenius law

\begin{equation}\label{eq_tau}
\tau=\tau_{0}{e^{\frac{E_{B}}{k_{B}T}}}
\end{equation}where $\tau_{0}$ directly depends on the material parameters ($K$, $M_{S}$, etc) and is of the order of $10^{-11} - 10^{-9} s$. Under this description, it therefore points out that the measuring time $\tau_{m}$ will be a key-point on determining the magnetic state of the system: if the measuring time is large in comparison with the characteristic reversal time of the particles, $\tau_{m} >> \tau$, then the particle moment will fluctuate rapidly from one well to the other in a paramagnetic-like fashion, i.e. the particle will be in the SPM state. However, if $\tau_{m} << \tau$, during the measuring time the particle moment will remain blocked along one anisotropy well, i.e. the particle will be in the BL state. Macroscopically, the SPM state is completely reversible upon temperature and field variations, whereas the BL one is characterized by its hysteretic behavior, proper of ferromagnets. The limit between both states is found at ${\tau_{m}}\equiv{\tau}$, and serves for the definition of the so-called \textit{blocking temperature} $T_{B}$, as illustrated in {\sc Figure}~\ref{FIG.3-tau_m} and obtained from Eq.(\ref{eq_tau})

\begin{equation}\label{eq_TB}
T_{B}=\frac{KV}{k_{B}\ln{(\tau_{m}/\tau_{0})}}
\end{equation}

As Eq.(\ref{eq_TB}) shows, $T_{B}$ depends both on the intrinsic particle parameters and on the external ones as the measuring time. Therefore, by varying the external influences of the systems (temperature, measuring time, applied field) we may tune its response. From the trends obtained we can extract information about the characteristics of the system, and so information for the design of technological applications.

Eq.(\ref{eq_TB}) highlights that the value of $\tau_{m}$ is decisive in determining $T_{B}$ and so the magnetic response of the system in a given time-scale (for example, for information storage-related applications, very large time scales have to be considered; however, for magnetic recording speed very short ones must be achieved). Since $\tau_{m}$ is determined by the experimental technique, its value has to be chosen with respect to the information and uses concerning our purposes. In this work we have focused on quasi static process, and so associate $T_{B}$ with the one obtained in dc-thermomagnetization measurements, in which the measuring time is very large large ($\tau_{m}\approx{100s}$) in comparison with the characteristic time $\tau_{0}$ of the particles. Other measurement techniques involving much shorter measuring times are associated to \textit{dynamic} measurements, not considered in this work\cite{dynamic_properties}. We have mainly focused on data obtained following the standard \textit{zero field cooling} (ZFC) and \textit{field cooling} (FC) protocols, in which the system is perturbed under a low magnetic field for measuring the evolution of its magnetization with temperature. Previous to describing such processes and its characteristics, we shall analyze the effect of the magnetic field on the magnetic properties of the single-domain nanoparticles.

\subsection{Field dependence}
\label{sec:FieldDependence}

When an external magnetic field is applied over the nanoparticles, it tries to align their magnetic moments along its direction. Therefore, except if applied perpendicularly to the easy anisotropy axis, it will favour the occupancy of one of the anisotropy energy wells over the other. The orientation of the magnetic moment of a particle \textit{i} is then governed by the competition among its uniaxial ($E_{A}$) and Zeeman ($E_{Z}$) energies

\begin{equation}\label{eq_Ea-Ez}
E^{i}=E^{(i)}_{A}+E^{(i)}_{Z}=-KV_{i}\left({\frac{{\vec{\mu}_{i}}\cdot{\hat{n}_{i}}}{\left|{\vec{\mu}_{i}}\right|}}\right)^2-{\vec{\mu}_{i}}\cdot{\vec{H}}
\end{equation}

The influence of the external field in the orientation of the magnetic supermoments, assuming coherent rotation of the atomic magnetic moments and the field applied at a certain angle $\theta_{0}$ with respect to the easy anisotropy axis, is known as the \textit{Stoner-Wohlfart model} \cite{stoner-wohlfart_1948} after the authors who first considered and solved this problem. They ignored thermal effects and so could solve it from minimal energy arguments, not taking into account time-dependence as related to temperature. In this introduction section, however, we are mainly interested in giving an overall description of the SPM features, as how the magnetic field will affect the orientation of the magnetic moments with temperature in relation to the anisotropy energy barrier and its influence on $T_{B}$. Therefore, we do not discuss now the orientation of the field at different angles and focus, for the sake of simplicity, on the simple case of the field applied parallel to the easy anisotropy axis ($\vec{H}\|{\hat{n}_{i}}$). Note that in this context of non-interacting particles, applying the magnetic field at a certain angle with respect to the easy axis is equivalent to consider only its projection along the axis. Since we are under the assumption of non-interacting and equivalent particles, we can apply single-particle considerations and simplify the \textit{i} subindex in Eq.(\ref{eq_Ea-Ez}), which taking into account Eq.(\ref{eq_mu}), reads

\begin{equation}\label{eq_EaEz-simplif}
E=-KV\cos^{2}{\theta}-M_{S}VH\cos{\theta}
\end{equation}

Eq.(\ref{eq_EaEz-simplif}) has two local energy minima at $\theta=0,\pi$ with values $E_{min}=-KV\pm{M_{S}VH}$, and a maximum a $\theta=\pi/2$ with value $E_{max}=KV(HM_{S}/2K)^{2}$. The $\theta=0$ value stands for the parallel orientation of the particle moment with respect to the magnetic field ($\uparrow\uparrow$), whereas the $\theta=\pi$ value stands for the antiparallel one ($\uparrow\downarrow$). This difference in the energy wells described by {\sc Figure}~\ref{FIG.2-NPmoment,eA-wells} corresponds therefore to different energy barriers depending on the orientation of the particle moment with respect to the applied field, which we shall call $E^{\uparrow\downarrow}_{B}$ and $E^{\uparrow\uparrow}_{B}$ for the antiparallel and parallel cases, respectively. Introducing the \textit{anisotropy field} of the particles as

\begin{equation}\label{eq_Ha}
H_{A}=\frac{2K}{M_{S}}
\end{equation}we calculate these energy barriers as the difference between the minima and maximum energies, obtaining

\begin{equation}\label{eq_EB-antiparal}
E^{\uparrow\downarrow}_{B}=KV\left({1-\frac{H}{H_{A}}}\right)^{2}
\end{equation}

and

\begin{equation}\label{eq_EB-paral}
E^{\uparrow\uparrow}_{B}=KV\left({1+\frac{H}{H_{A}}}\right)^{2}
\end{equation}

\begin{figure}[ht!]
\begin{center} 
\includegraphics[draft=false,width=1.00\textwidth]{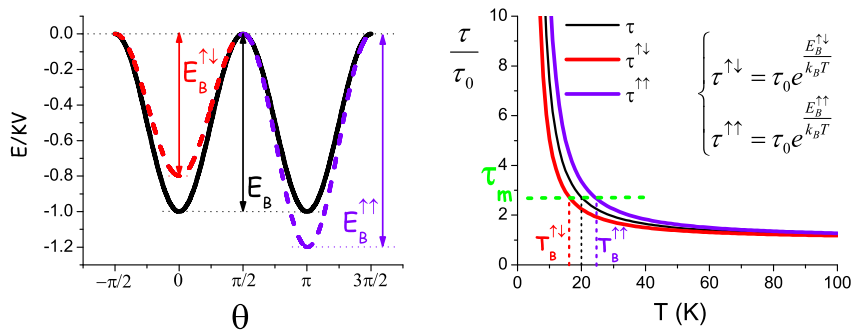}
\end{center} 
\caption{Anisotropy energy wells (left panel) and relaxation time (right panel) of the particles as influenced by the magnetic field.}\label{FIG.4-eAeZ-wells}
\end{figure}

The difference in the height of the energy barriers indicates also a change in the characteristic relaxation time of the particles, since it depends now on the relative orientation of the magnetic dipoles with respect to the field: particles antiparallel-oriented with respect to the field have a smaller energy barrier in comparison to the solely-anisotropy one and so a smaller thermal energy is enough to overcome it, whereas the parallel-oriented particles are now confined into a deeper anisotropy well and so a higher thermal energy is necessary to promote the jumping of its magnetic moments. This influence of the magnetic field on the energy wells of {\sc Figure}~\ref{FIG.2-NPmoment,eA-wells} is shown in {\sc Figure}~\ref{FIG.4-eAeZ-wells} (left panel), as well as its implications into the relaxation time of the particles (right panel).

{\sc Figure}~\ref{FIG.4-eAeZ-wells} illustrates the importance of the strength of the applied magnetic field on determining the magnetic properties of SPM systems, and also serves as a definition of \textit{small field} as referred to magnetic nanoparticles in comparison with their anisotropy field: the ratio $H/H_{A}$ must be as smaller as possible so that the system keeps as closer as possible to the ideal SPM conditions.

\subsection{Thermomagnetic measurements}
\label{sec:ThermomagneticMeasurements}

\begin{figure}[ht!]
\begin{center} 
\includegraphics[draft=false,width=0.65\textwidth]{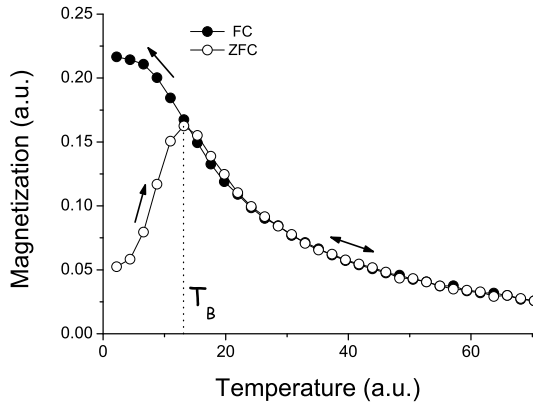}
\end{center} 
\caption{Typical ZFC and FC magnetization curves \textit{vs} temperature. Vertical dotted line stands for the maximum of the ZFC curve, usually associated to $T_{B}$.}\label{FIG.5-zfc-fc}
\end{figure}

Once we have gone through how a magnetic field $\vec{H}$ influences the properties of the particles we can undertake the description of the \textit{ZFC} and \textit{FC} measurements. In both processes the temperature evolution of the total magnetization of the system is recorded following different thermomagnetic histories, and it is this different history what highlights reversibility (no-hysteresis) and irreversibility (hysteresis) for differentiating the anhysteretic SPM state from the hysteretic BL state\cite{evaluating_hysteresis}. In a ZFC process, the system is first cooled down in zero field until a very low T is reached, and afterwards a small field is applied and the magnetization is recorded while heating the sample up. The FC curve is obtained by measuring the magnetization of the sample while cooling under low magnetic field (same field strength for both ZFC and FC processes, and low to minimize the disturbance of the system). Typical ZFC/FC curves are shown in {\sc Figure}~\ref{FIG.5-zfc-fc}.

The ZFC and FC curves shown in {\sc Figure}~\ref{FIG.5-zfc-fc} display the usual features found in SPM systems: i) both curves coincide at high temperatures in a PM-like dependence; ii) with decreasing T both curves grow until a certain temperature range is reached where the curves start to diverge, the FC curve still growing although at a lower rate while the ZFC exhibits a maximum and decreases below it. This maximum in the ZFC curve is generally associated in the literature to $T_{B}$ \cite{review-kechrakos_JNN}, as indicated in the figure, since such maximum roughly differentiates two main temperature regimes: a high-temperature one where both curves essentially coincide and exhibit a PM-like temperature dependence, from a low-temperature regime where both curves clearly diverge. However, a detailed view of the curves reveals that a $1/T$ PM-like decrease right above $T_{B}$ is not observed in the ZFC curve, and a slight difference between the ZFC and FC one is perceived. These features indicate that a true SPM behavior is not exhibited right above the maximum, but only at higher temperatures the ZFC curve perfectly overlaps the FC one and exhibits a well-defined PM-like temperature dependence. The reason why the ideal SPM behavior is not observed right at $T > T_{B}$ is found on the inverse of the relaxation time $\tau$, which gives the probability of the particle to overcome the anisotropy energy barrier along the temperature

\begin{equation}\label{eq_inv_tau}
p(T)=\tau^{-1}_{0}e^{\frac{-E_{B}}{k_{B}T}}
\end{equation}

Eq.(\ref{eq_inv_tau}) considering the different energy barriers $E^{\uparrow\downarrow}_{B}$ and $E^{\uparrow\uparrow}_{B}$ and so different relaxation times (see {\sc Figure}~\ref{FIG.4-eAeZ-wells}), indicate that SPM behavior will be only observed (ideally) above $T^{\uparrow\uparrow}_{B}$, which above the maximum of the curve. Therefore, the features displayed in {\sc Figure}~\ref{FIG.4-eAeZ-wells} concerning the different height of the anisotropy wells as influenced by the external field, must be taken into account too when analyzing the physical trend followed in the ZFC magnetization curve. The initial state of the ZFC process starts with no net magnetization after cooling in zero field. If naming \textit{parallel} particles those with $E_{Z} < 0$, and \textit{antiparallel} particles those with $E_{Z} > 0$, then when the field is applied parallel particles will rapidly align with the field, while the antiparallel ones will progressively overcome the energy barrier with the increasing thermal energy and also align with the field. This process leads to a continuous increase of the magnetization, as illustrated in {\sc Figure}~\ref{FIG.5-zfc-fc}, until the thermal energy overcomes $E^{\uparrow\downarrow}_{B}$, and so no longer reversal magnetization takes place. This is not so simple however, since the thermal energy does not induce the alignment of more particles, but counterbalances their orientation, inducing thermal fluctuations. As the thermal energy is now comparable to the energy of the deeper anisotropy well, inverse reversal mechanism overcoming has a higher appreciable probability to occur. This frame stands so for the expected further decrease in the magnetization, although not yet accomplished full PM-like behavior. In fact, pure PM-like behavior would only be observed for $k_{B}T >> E_{B}$. It is important to emphasize these aspects when dealing with the magnetic properties of SPM systems, remarking that $T_{B}$ defines only a characteristic temperature value, of enormous interest for characterizing the system but not equivalent to a phase transition\cite{Tb_PT}. Because of the above reasons, some authors do not associate directly the maximum in the curves to $T_{B}$ and prefer a different definition \cite{review-knobel_JNN}. Although we understand such discrepancy and share the necessity of finding a more precise formalism for relating the shape of the ZFC curves to the particles' characteristics, we associate in this work the maximum of the ZFC curve with $T_{B}$ because it is the usually preferred when dealing with experimental data. 

\begin{figure}[ht!]
\begin{center} 
\includegraphics[draft=false,width=0.80\textwidth]{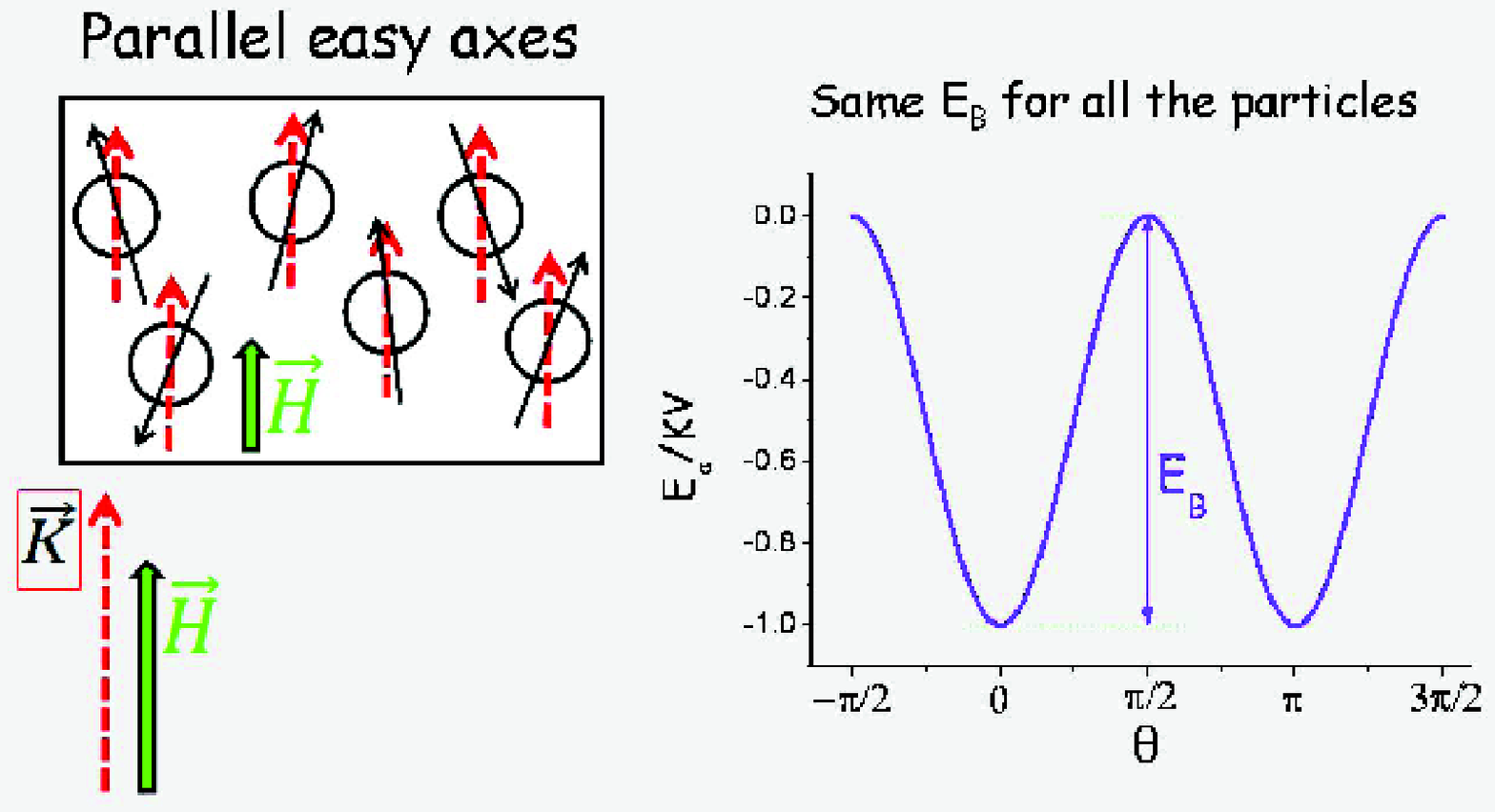}
\end{center} 
\caption{Schematic drawing of the ideal simplest model of non-interacting and parallel aligned easy axes along the applied field.}\label{FIG.6-aligned_axes}
\end{figure}

It is important to remind here that all the properties described up to now correspond to the simplest single-particle scenario, where all the particles have been treated as equivalent to each other because of been non-interacting and with parallel-aligned easy anisotropy axes, as illustrated in {\sc Figure}~\ref{FIG.6-aligned_axes}.

\begin{figure}[ht!]
\begin{center} 
\includegraphics[draft=false,width=0.80\textwidth]{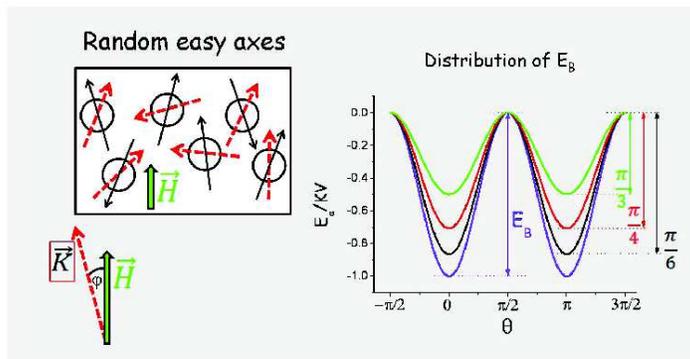}
\end{center} 
\caption{Schematic drawing of the usual random easy axes distribution, and the modulation of the energy barrier height along the field direction.}\label{FIG.7-random_axes}
\end{figure}

However, real systems are usually characterized by a random distribution of the anisotropy easy axes, what would be equivalent to a distribution of the heights of the energy barriers along the field direction, as schematized in {\sc Figure}~\ref{FIG.7-random_axes}, giving place to a extraordinarily complex scenario. Moreover, this situation becomes even more complex in real systems, for one of the main problems concerning the magnetic properties of SPM systems is the role played by interparticle dipole-dipole interactions, which are long-range and anisotropic, resulting so in a complex interplay with the anisotropy barriers. Low-interacting conditions can be described by mean-field approximations in which the single-particle barriers are slightly modified by the interaction energy. However, strongly-interacting conditions ruled by collective effects cannot be accounted by that approach, and so the use of computational techniques becomes an indispensable tool for treating systems with so many freedom degrees. Computational techniques allow us to deal with perfectly controlled systems and the exact treatment of the interactions among particles. There are two main computational approaches for dealing with the magnetic properties of interacting nanoparticle systems, namely the \textit{Monte Carlo} (MC) and \textit{Langevin Dynamics} (LD) methods. Both methods are complementary for the study of a nanoparticle system: MC simulations are very adequate to treat long-time (static) magnetic properties in complex interacting systems \cite{garcia-otero_prl} but do not have associated a physical time; LD methods, on the contrary, are very precise for modeling the dynamics of the magnetic moments \cite{usadel_prb2006} but cannot treat static processes.\cite{LD-time_MC} As mentioned above we have focused our description of the SPM phenomena by means of its features throughout static measurements; next we introduce the MC technique in the context of studying complex SPM systems as the ones described here.

Finally, it is worthy to recall again the several simplifications assumed in this introduction to \textit{superparamagnetism}, where we have considered very simple and ideal characteristics for the particles. In real systems there are always several dispersive -often uncontrolled- characteristics (inhomogeneities in the particles' composition; temperature-dependent $K$ and/or $M_{S}$; size/anisotropy distribution; etc), characteristics all that confer additional uncontrolled degrees of freedom to the already very complex system, and so make very hard to interpret the magnetic measurements in order to characterize their properties. That is the reason why we had to focus on a very simple scenario which, on the other side, it is already complex enough to make absolutely necessary the use of computational techniques for its study.

\section{Monte Carlo method}
\label{sec:MonteCarloMethod}
\begin{figure}[ht!]
\begin{center} 
\includegraphics[draft=false,width=1.00\textwidth]{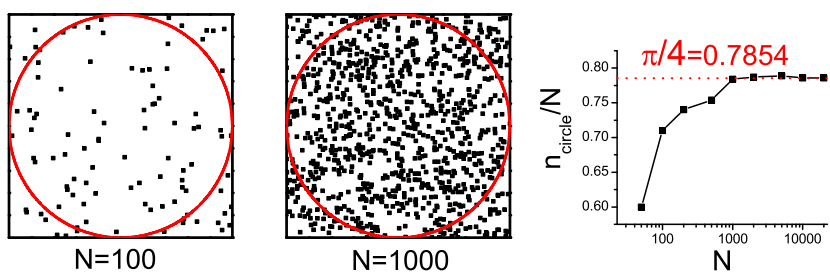}
\end{center} 
\caption{Use of random events to calculate the value of $\pi$ from the area of a circle, and the importance of the amount of events in the precision in the result.}\label{FIG.8-example-circle_MC}
\end{figure}

\textit{Monte Carlo (MC) methods} are a type of numerical simulation techniques based on the generation of random numbers \cite{binder}. MC methods are utilized to solve complex problems with large freedom degrees: the features of a particular problem are represented by probabilities, and the MC technique consists on generating large amounts of random numbers and counting the fraction of them obeying some conditions. The way of counting and the conditions imposed define the \textit{numerical algorithm}. A simple example to illustrate the functioning of a MC method is the calculus of $\pi$ from the area of a circle. If placing a circle of radius $r$ into a square of side $2r$ and randomly generating $N$-points into the square, the fraction of them laying inside the circle ($n_{circle}$) will be equal to the proportion between areas, and so it is easy to obtain $\pi=4\frac{n_{circle}}{N}$. The MC calculation of the area will be more precise the larger the amount of points (events) generated, as illustrated in {\sc Figure}~\ref{FIG.8-example-circle_MC}.

The two drawings in {\sc Figure}~\ref{FIG.8-example-circle_MC} stand for two examples of random generation of events (N=100 and N=1000) into the square with the circle held inside. The graph shows the fraction between points laying in each geometrical figure as a function of the amount of random events generated, pointing out that the precision to determine $\pi$ from the random points areal ratio grows rapidly with N.

\subsection{Metropolis algorithm}
\label{sec:SimulationOfSuperparamagneticSystemsMetropolisAlgorithm}

We aim in this work to describe how to use a MC method to simulate the magnetic properties of a magnetic nanoparticle system with a large amount of freedom degrees as described in the previous {\sc Section}~\ref{sec:Superparamagnetism}. Specifically, we want to know how the orientation of the magnetic moments of the particles evolves as a function of different parameters (temperature, applied magnetic field, etc), i.e. how they behave as a function of the different energies involved. To simulate such processes we impose to the system some known conditions and determine its configuration from a random generation of events as evaluated under those conditions. The processes we want to simulate are essentially quasi static, and so the conditions ruling the system can be based on minimum energy arguments on the following manner: i) the energy of the system under certain conditions is evaluated, ii) an external parameter is varied and the energy is reevaluated, and iii) the difference in energy is used to construct a probability function, and the change of configuration of the system is accepted or not from the random generation of events applied to such probability under a given algorithm.

It turns out that this problem is much more complex than calculating $\pi$ from the ratio between the areas of the circle and square as described above, where the randomly generated events are equally probable and so the algorithm for solving the problem is just to count 1 if the points lay inside the circle and 0 otherwise. If we apply the same procedure to simulate the orientation of the magnetic moments with temperature, i.e. if we randomly generate new possible orientations and evaluate its feasibility to occur, we will find that most of the trials are highly improbable and so rejected, and only those with energy comparable to the previous state will have some chance of being accepted. For example, for simulating the new possible orientation of a particle' magnetic moment, initially at an angle $\theta_{i}$ (Eq.(\ref{eq_EaEz-simplif})), the new trial configuration $\theta_{f}$ can be chosen totally at random, unrelated to $\theta_{i}$, or by considering a slight variation after the actual configuration, so that $\theta_{f}=\theta_{i}+\delta{\theta}$, with $\delta{\theta}$ small. In the first case many trials will be very unfavorable and so rejected, while for the latter a higher acceptance ratio is expected. It becomes therefore crucial, in order to avoid the wasting of computational time and for optimizing the simulation, to be able to select the new trial configurations among the most likely probable paths. This can be done if considering a Markov chain of events (configuration of one state depends only on the previous one), with the trial state being close in energy to the current one.

The key-points for treating the present problem are, therefore, i) the selection of the trial configurations in an efficient way, and ii) the choosing of an adequate implementation of the change from the initial state with energy $E_{i}$ to the trial next state with energy $E_{f}$. Assuming classical Boltzmann distribution, the probability of a magnetic moment to have energy $E$ at a temperature $T$ is proportional to $exp(-E/k_{B}T)$, i.e. $p(E)\propto{e^{-E/k_{B}T}}$. Consequently, if considering the orientation of the magnetic moments to be markovian the evolution from state $E_{i}$ to state $E_{f}$ will be proportional to the rate between final and initial states probabilities, $r_{i\rightarrow{f}}=p(E_{f})/p(E_{i})=e^{-\Delta{E}/k_{B}T}$, with $\Delta{E}=\left({E_{f}-E_{i}}\right)$. This way of choosing the possible next configuration of the system as being energetically close to the actual one is named \textit{importance sampling}, and is based on the \textit{detailed balance} reversibility condition. This approach works very well for describing quasi-static thermodynamic processes, as intended in this work, although much care has to be taken if dealing with dynamic properties. For further details about this topic, see for example: O. Iglesias \textit{Doctoral Thesis} \cite{iglesias_tesi}, Chapter 5.

The motion of the magnetic moment of a nanoparticle from the initial state with energy $E_{i}$ to the final state with energy $E_{f}$ is often described by means of the \textit{Metropolis algorithm} \cite{metropolis_algorithm}: if $\Delta{E}<0$ (the new configuration is more stable energetically), the move to the new state is accepted and its energy changes to $E_{f}$, whereas if $\Delta{E}>0$ (new configuration more unstable), although small, the move has still some probability $e^{-\Delta{E}/k_{B}T}$ to occur\cite{other_algorithms}. To compute this probability a random number $n$ with value between 0 and 1 is generated and, if $n < r_{i\rightarrow{f}}$ the new conformation of the system is accepted and so it has now energy $E_{f}$, while if $n < r_{i\rightarrow{f}}$ it is rejected and the energy remains still $E_{i}$. The Metropolis algorithm for the probability of a $\Delta{E}$ configuration change is expressed as 

\begin{equation}\label{eq_metropolis}
min\left[{1,e^{-\Delta{E}/k_{B}T}}\right]
\end{equation}

The Metropolis MC method can be used to simulate the evolution of the magnetization of a system of magnetic nanoparticles as a function of different parameters. We describe here a MC method based in this procedure (see for example Refs. \cite{serantes_phys.stat.sol.a, serantes_porto, serantes_prb2009, serantes_j.phys.d, serantes_jnn, serantes_prb2010, serantes_jap2010}): the simulated system consists on an assembly of N-particles contained inside a unit cubic cell (side $L$), which is replicated by using periodic boundary conditions in order to resemble a large and homogeneous system. The simulations are always done in two parts: in the first one the spatial distribution of the particles is set, and in the second part the particles are characterized by their attributes (size, anisotropy, magnetic moment) and the evolution of the system is simulated as a function of the desired parameters. The positions set in the first part are kept fixed during the simulation of the magnetic properties. Next we give a brief description of the generalities of the MC method.

\subsection{Spatial arrangement}
\label{sec:SpatialArrangement}

The procedure used to achieve the spatial distribution of the particles varies depending on the type of system we want to simulate: for ordered structures (chains of particles, crystalline structures, etc) the particles are directly placed into the unit cubic cell under the desired structure, whereas for setting the spatial arrangement of liquid-like systems (e.g. a frozen ferrofluid) a relaxation algorithm has to be used.

\begin{figure}[ht!]
\begin{center} 
\includegraphics[draft=false,width=1.00\textwidth]{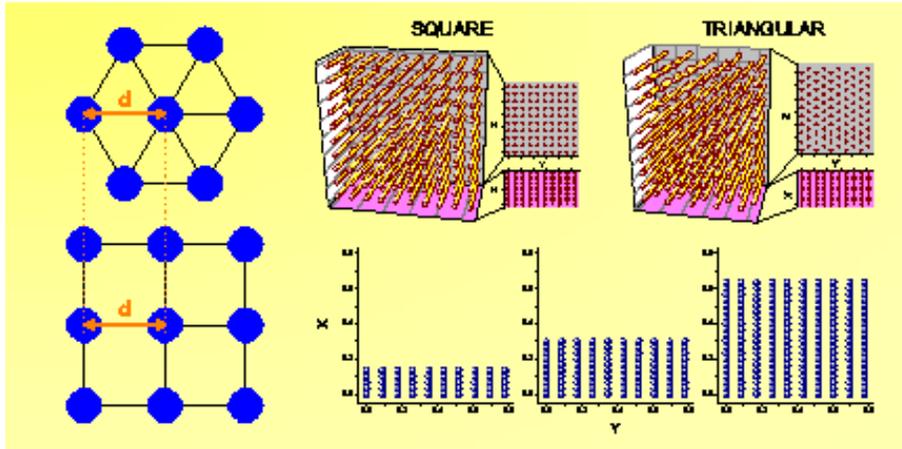}
\end{center} 
\caption{Scheme of parallel-aligned chain-like structures hexagonally and squared distributed, and different lengths.}\label{FIG.9-debuxo-estructuras}
\end{figure}

In {\sc Figure}~\ref{FIG.9-debuxo-estructuras} we show some chain-like structures obtained by directly placing the particles into desired regular positions. They resemble one-dimensional columnar parallel chains of magnetic nanoparticles under different spatial arrangement (square, hexagonal) and different lengths along the X-axis. The study of such type of structures is at the center of much research nowadays for the basic study of the competition between the enhanced anisotropy and magnetostatic interactions \cite{lavin_jap2010, lim_jap2010, franco_prb2008}. The magnetic properties of such chain-like systems exhibit a good analogy to the behavior of ferromagnetic nanowires \cite{prida_pssa2009}.

\begin{figure}[ht!]
\begin{center} 
\includegraphics[draft=false,width=1.00\textwidth]{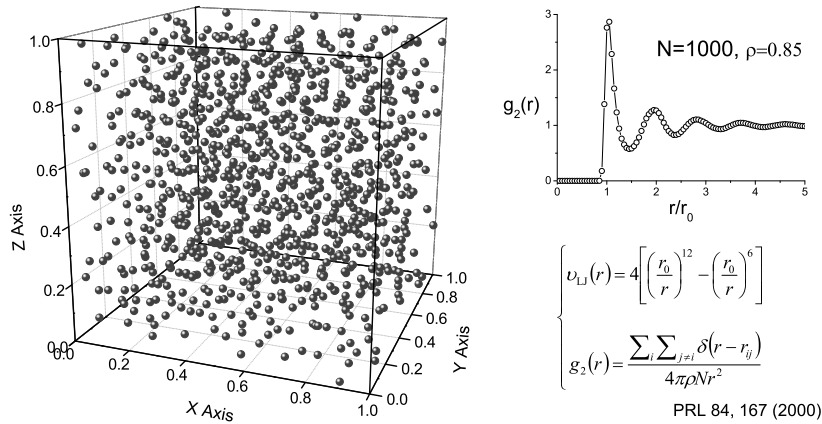}
\end{center} 
\caption{Liquid-like distribution of N=1000 particles into a unit cubic cell and correlation function $g_{2}(r)$. Both the correlation function and Lennard-Jones pair potential are also indicated.}\label{FIG.10-debuxo-liquido}
\end{figure}

For simulating disordered systems as ferrofluids with liquid-like structure \cite{luo_prl1991} the positions of the particles are not directly generated and so we use a Lennard-Jones pair potential ($\upsilon_{LJ}$) to distribute the particles. During the simulation the particles can move freely in space, but their trial positions are markovian-linked to the actual one and so the liquid-like structure is more quickly obtained. An example of liquid-like structure is shown in {\sc Figure}~\ref{FIG.10-debuxo-liquido}, together with the corresponding correlation function.

For treating the long-range dipolar interactions the Ewald summation is used as in Ref. \cite{garcia-otero_prl}, considering for simulating long and homogeneous systems periodic boundary conditions over repetition of the unit cubic cell. The positions of the particles and their relative interdistances are calculated and storage for the next part of the simulations, in which remain constant throughout.

\subsection{Superparamagnetic (SPM) properties}
\label{sec:SuperparamagneticProperties}

Once the spatial distribution of the particles is achieved the next step is to characterize them with their main physical properties, namely volume, magnetic moment and magnetic anisotropy. Following the model described in {\sc Section}~\ref{sec:Superparamagnetism}, magnetic anisotropy is considered of uniaxial type, and both magnetic moment and anisotropy are assumed to be proportional to the particle volume, and so the important parameters to characterize the particles are their volume and the orientation of the magnetic moment and easy-anisotropy axis.

The volume is taken into account by means of the related sample concentration of the system $c$, preferable to determine experimentally and so better to compare with experimental results. For the sake of simplicity we assume the same monodisperse system as in the previous section, in order to have the less uncontrolled parameter-dispersity as possible. If defining $c$ as the fraction of the volume occupied by all the particles ($\sum^{N}_{i}{V_{i}}$) over the total system volume ($V_{T}=L^{3}$), then the relationship between particle size and volume sample concentration is

\begin{equation}\label{eq_sample-concentration}
c=\frac{\sum^{N}_{i}{V_{i}}}{L^{3}}\equiv{\frac{NV}{L^{3}}}
\end{equation}

The orientation of the anisotropy easy axes of the particles is a parameter that remains fixed along the whole simulation process, and hence its value must be carefully selected due to its strong influence on the magnetic properties of the system. For studying SPM properties we have assumed a random easy-axes distribution as schematized in {\sc Figure}~\ref{FIG.7-random_axes}, constraint that works quite well for describing real systems as frozen ferrofluids or solid matrices \cite{hoppe_jpcc}. Once the easy axes are placed, the simulation of the magnetic evolution of the magnetic moments of the particles can start, whose initial orientation can also be randomly distributed. The magnetic properties of the particles are contained on their magnetic moment $\vec{\mu}$ and magnetic anisotropy $\vec{K}$, under the \textit{temperature/magnetic field} constraints applied to the system, which determine its energy. In real systems, the evolution between an initial state with energy $E_{i}$ and a final state with energy $E_{f}$ occurs at a certain time interval, in which the event (reorientation of the magnetic moments in this case) can be described by a probability distribution function. These time-dependent processes can be simulated by giving the system a certain amount of chances to occur, i.e. \textit{attempts} to change the configuration. These attempts define the \textit{computational time} and are called \textit{Monte Carlo steps}. The MC step constitutes, therefore, the computational equivalence to real time units. In our simulations one MC step is defined as $N$ trials given to a system of $N$-particles to change its configuration.

\begin{figure}[ht!]
\begin{center} 
\includegraphics[draft=false,width=0.85\textwidth]{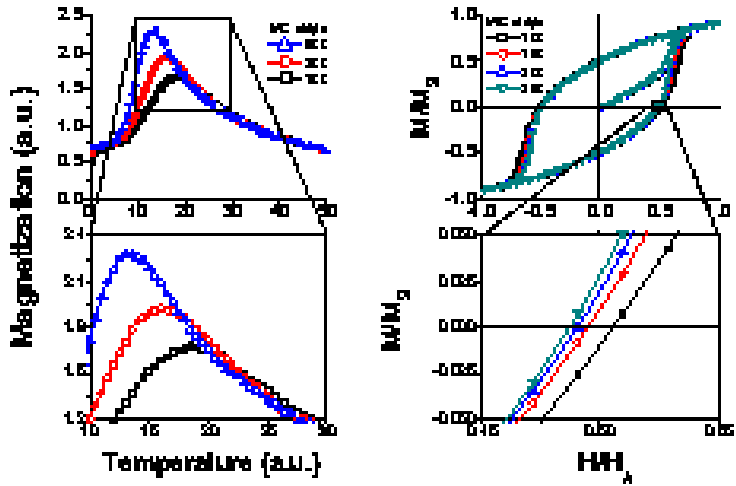}
\end{center} 
\caption{Left panel: ZFC processes simulated for a fixed temperature interval variation, but different MC steps (100, 200, and 500). Right panel: M(H) simulated processes simulated for a fixed magnetic field interval variation, but different MC steps (100, 150, 200, and 250). In both cases, top panel shows the complete processes, while bottom panel shows a magnification of the more remarkable aspects of the simulation (maximum of the ZFC and coercive field, respectively).}\label{FIG.11-MCsteps}
\end{figure}

The simulation of a physical process consists in varying a desired magnitude (temperature, magnetic field, etc) under a certain protocol and evaluating the energy in the new state after a certain number of MC steps, accepting or neglecting the new configuration under the chosen algorithm. For example, to simulate a ZFC process in a system of $N$-particles the system is cooled down from a high temperature in regular temperature intervals every a certain fix amount of MC steps in zero field down to a very low temperature. In every MC step, one particle is selected at random and a new orientation of its magnetic moment is generated and accepted under the Metropolis algorithm $min\left[{1,e^{-\Delta{E}/k_{B}T}}\right]$. This is repeated $N$-times in each MC step. Once the very low temperature is reached, a small magnetic field is applied and the process continues now while heating the sample\cite{thermalization}. The same procedure applies for simulating the magnetization \textit{vs} magnetic field M(H) curve, just being different the parameter to vary after a certain amount of MC steps. A M(H) curve is simulated by initially cooling the system down to the desired temperature (in zero field for our simulations), and once it is reached, a small field is applied and increased in regular intervals of field/MC steps up to a high field $H_{max}$. Then, the field is decreased in the same manner until $-H_{max}$ is reached, and finally increased again until reaching once more $H_{max}$ and having completed the cycle. To illustrate the features of the MC steps resembling real time units, we show in {\sc Figure}~\ref{FIG.11-MCsteps} some ZFC (left panel) and M(H) (left panel) curves for fixed temperature and magnetic field variations, but different MC steps, corresponding to a system of non-interacting particles as that shown in {\sc Figure}~\ref{FIG.10-debuxo-liquido}.

Left panel of {\sc Figure}~\ref{FIG.11-MCsteps} shows that the curves peak at lower temperatures for larger amount of MC steps, what reproduces the physical behavior described by Eq.(\ref{eq_TB}): if relating the amount of MC steps with the experimental measuring time $\tau_{m}$, then the longer is the time interval (amount of MC steps), the smaller it $T_{B}$. The behavior observed in the M(H) curves of {\sc Figure}~\ref{FIG.11-MCsteps} (right panel) indicates a decrease in the coercive field for larger time intervals (higher amount of MC steps), also as expected since the FM-like behavior represented by the area in the M(H) curves is time-dependent, as discussed in {\sc Section}~\ref{sec:Superparamagnetism}, and tends to disappear for very large times (MC steps). However, it is important to indicate here that, although the physical tendency coincides with the expected in both kind of simulations, there is not a well-established relationship between MC steps and real time units and some scaling must be taken in this aim \cite{nowak-prl2000, cheng-prl2006}. As mentioned above, time-dependent processes in systems of magnetic nanoparticles are described better by means of the Landau-Lifshitz-Gilbert dynamic equation \cite{usadel_prb2006}.

\subsection{Trial \textit{computational time} steps}
\label{sec:TrialTextitComputationalTimeSteps}

\begin{figure}[ht!]
\begin{center} 
\includegraphics[draft=false,width=0.30\textwidth]{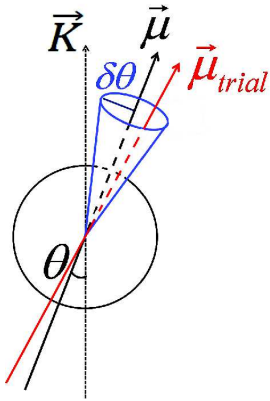}
\end{center} 
\caption{Schematic drawing of the $\delta{\theta}$-cone used to generate the new trial orientation $\vec{\mu}_{trial}$.}\label{FIG.12-THETA_JUMP}
\end{figure}

It turns out from the above description of the MC technique that the choosing of the \textit{trial} new configuration with energy $E_{f}$ is of primordial importance: its value determines the acceptation rate of the algorithm and so the velocity and feasibility of the simulations. In our simulations we use the so-called \textit{solid angle restriction scheme} \cite{pereira-nunes_pre2004}: the new trial orientation $\vec{\mu}_{trial}$ is randomly generated inside a cone of angle $\delta{\theta}$ around the current orientation $\vec{\mu}$. {\sc Figure}~\ref{FIG.12-THETA_JUMP} illustrates this choice of the new trial orientation inside a cone of angle $\delta{\theta}$ around the current orientation of the magnetic moment.\cite{other_MCtrials}

The $\delta{\theta}$-value directly rules the speed of the magnetic moments' movement, and so its magnitude must be carefully selected in order to ensure the adequacy of the simulations to resemble physical processes: if $\delta{\theta}$ is too small the system will evolve very slowly to the quasi-equilibrium configuration, and we could be unable to resemble the physical process (too many MC steps would be necessary). If $\delta{\theta}$ is too large, the system can relax too fast and make the features we want to study to become unobservable. In {\sc Figure}~\ref{FIG.13-THETA_cte,THETA-t}(a), we show the ZFC curves of the same system obtained for different values of $\delta{\theta}$. It is observed a tendency similar to that displayed in {\sc Figure}~\ref{FIG.11-MCsteps}, with the curves exhibiting a larger peak at decreasing temperature for larger $\delta{\theta}$-values. These results support our arguments on the functioning of the trial steps as \textit{computational time} for reproducing real time-dependent processes.

\begin{figure}[ht!]
\begin{center} 
\includegraphics[draft=false,width=0.85\textwidth]{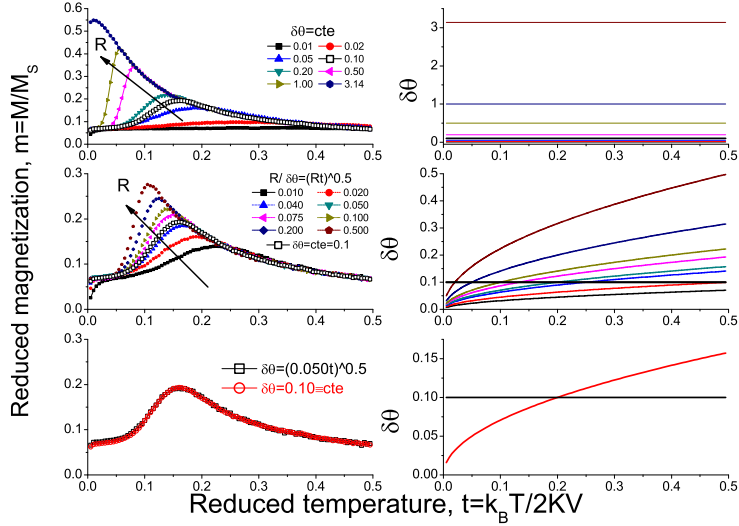}
\end{center} 
\caption{ZFC processes as a function of $\delta{\theta}$. (a) corresponds to different $\delta{\theta}$-values constant with temperature, (b) shows different temperature-dependence for $\delta{\theta}$, and in (c) we compare the $\delta{\theta}$=0.1=cte case, with being temperature-dependent as $\delta{\theta}=(0.05t)^{1/2}$. In all cases, panels on the right show the temperature evolution of $\delta{\theta}$ and the $\delta{\theta}$=0.1=cte case is shown for the comparison.}\label{FIG.13-THETA_cte,THETA-t}
\end{figure}

The results plotted in {\sc Figure}~\ref{FIG.13-THETA_cte,THETA-t} indicate a strong dependence of the simulated results on $\delta{\theta}$, emphasizing so that much care must be taken if trying to extract time-related information about the system from the simulations: those depend not only on the artificial \textit{MC step}, but also on the trial angle chosen. As pointed out at the beginning of this subsection, the Monte Carlo technique is very successful for analyzing quasi-static processes, and so we have focused in this work on the more reliable aspects of the simulations, i.e., quasi-static analysis of complex processes with large freedom degrees.

{\sc Figure}~\ref{FIG.13-THETA_cte,THETA-t} displays also some simulations in which the $\delta{\theta}$-value is considered to be temperature-dependent, in the same way as described in Refs. \cite{nowak-prl2000, cheng-prl2006}. The reason for including such dependence is not related to any effort concerned to use realistic times from the MC steps or to reach more accurate time-dependence results (Figs. \ref{FIG.11-MCsteps} and \ref{FIG.13-THETA_cte,THETA-t}(a) illustrate that it constitutes a very complex task). Instead, our motivation was simply to provide the simulations with a more realistic character: it seems to us more physical to make the thermal fluctuations to be larger at larger temperatures. Our argument is based on the fact that the analogy between the time-dependence of real experiments and simulations is introduced by means of \textit{attempts} to change the configuration, which are generated randomly into the cone of angle $\delta{\theta}$. Under this assumption, the \textit{movement} of the magnetic moments is resembled by giving the system a certain amount of MC steps to change its configuration.

In a real system, the magnetic moments fluctuate because of the thermal energy and consequently fluctuations are reduced the smaller the temperature. In order to reproduce this characteristic in the simulations, it appears very reasonable to us to include a temperature-dependence in the value of $\delta{\theta}$, since it is the tool used to resemble the thermal-flucuations found in real systems. It is worth no note here that although there is no intention of analyzing time-dependence in our simulations, however the temperature-dependence expression used has been intensively analyzed and discussed in such a context by Nowak et al. \cite{nowak-prl2000} and Cheng et al. \cite{cheng-prl2006}. The temperature dependence of $\delta{\theta}$ can be, following Refs. \cite{nowak-prl2000, cheng-prl2006} and in usual computational reduced temperature units\cite{serantes_phys.stat.sol.a, serantes_porto, serantes_prb2009, serantes_j.phys.d, serantes_jnn, serantes_prb2010, serantes_jap2010} $t=k_{B}T/2KV$, written as $\delta{\theta}=(Ct)^{1/2}$, with C a constant value proportional to the particle inner characteristics (size, magnetic moment and anisotropy) and to the dynamics of the system (gyromagnetic ratio, damping parameter, measuring time). The results plotted in {\sc Figure}~\ref{FIG.13-THETA_cte,THETA-t}(b) show that for values of $C$ of the order of 0.050, not large variations are considered in comparison with the $\delta{\theta}=0.10$ case, as shown in {\sc Figure}~\ref{FIG.13-THETA_cte,THETA-t}. We arbitrarily choose the $\delta{\theta}=0.050t$ dependence value for taking into account temperature-dependence of $\delta{\theta}$. As a general rule, in our simulations we have considered the $\delta{\theta}=0.10$ trial cone, and only very recently the temperature dependence of $\delta{\theta}$ has been considered. For every calculation described on the next sections the way of choosing $\delta{\theta}$ will be properly described.

\section{Summary}
We have introduced the basic features of the so-called \textit{superparamagnetism} as the paramagnetic-like temperature dependence that occurs in single-domain magnetic systems above a characteristic \textit{blocking temperature}, $T_{B}$. The special SPM phenomena arises from the interplay between the large magnetic \textit{supermoment} arising from the coherent rotation of the atomic magnetic moments in single-domain entities, and the magnetic anisotropy resultant from the crystalline, shape, etc, contributions. The presence of a magnetic anisotropy defines a preferential orientation direction which plays a relevant energy term when the temperature is comparable to $T_{B}$. The direction of this anisotropy in relation to the applied magnetic field defines the response of the magnetic supermoment of the particle with temperature. In real systems, the characteristic parameters (mainly anisotropy and magnetic moment) vary from particle to particle due to the experimental difficulties to synthesize particles with perfectly controlled characteristics and to place them regularly (both the spatial position and the anisotropy orientation). In addition, real systems are also subjected to other effects as the interparticle magnetic dipolar interaction, adding more uncontrolled parameter degrees to the already \textit{per se} very complex theoretical scenario. In this context, we introduced the \textit{Monte Carlo} technique based on the \textit{Metropolis algorithm} as a powerful tool for the study of SPM properties.

\paragraph{Acknowledgements}
\label{sec:Acknowledgements}
We thank the Xunta de Galicia for financial support (Project INCITE 08PXIB236052PR), and the Centro de Supercomputaci\'{o}n de Galicia for the computing facilities.


\begin{thebibliography}{46}
\bibitem{dynamic_properties}For a recent review on the dynamic properties of SPM systems see for example the work by P.E. Jonsson\cite{review-jonsson}; other measurement techniques associated to different time scales are analyzed in Ref.\cite{dormann_fiorani_tronc}.
\bibitem{evaluating_hysteresis}Although for evaluating magnetic hysteresis measurements of M(H) cycles could seem a better choice, they are not appropriate for determining the value of $T_{B}$ since it would be necessary a large amount of temperatures where M(H) to be recorded. On the contrary, M(T) measurements give a continuous temperature range to more precisely obtain $T_{B}$ from a single curve.
\bibitem{Tb_PT}Other consideration related to the meaning of $T_{B}$ concerns the question intensively discussed in the literature about whether strong dipolar coupling among the particles could lead to a phase-transition character at $T_{B}$\cite{garcia-otero_prl,serantes_jnn}. However, this is not the present case since we are dealing with a non-interacting system.
\bibitem{thermalization}A thermalization is applied before any thermomagnetic simulation (ZFC, FC, magnetization vs. field) in order to rapidly reach a stable configuration. The reason for doing this is to reduce the initial instability of the system right after randomly generating the orientation of the magnetic moments and easy-anisotropy axes, which would take a vary large amount of MC steps to stabilize during the simulation process. Instead, if the system is initiated at a very high temperature and cooled in large temperature jumps down to the starting process temperature the system relax rapidly and reaches a energetically stable configuration.
\end{thebibliography}
\end{document}